\documentstyle[11pt,newpasp,twoside]{article}
\def\edcomment#1{\iffalse\marginpar{\raggedright\sl#1\/}\else\relax\fi}
\reversemarginpar

\begin{document}
\title{Pulsar Glitch Behaviour 
and AXPs, SGRs and DTNs}
\author{M. Ali Alpar}
\affil{Sabanc{\i} University, Orhanl{\i}, Tuzla 81474, 
Istanbul, Turkey}

\begin{abstract}
Large pulsar glitches seem to be common to all radio pulsars and to 
exhibit a universal behaviour connecting the rate of occurrence, 
event size and interglitch relaxation that can be explained 
if the glitches are due to angular momentum exchange in the neutron 
star. This has implications for the energy dissipation rate. 
The large timing excursions observed 
in SGRs are not similar to the pulsar glitches. 
The one glitch observed in an AXP is very similar to pulsar 
glitches and demonstrates that typical neutron star glitch response occurs 
when the external torque is constant for an extended period of time. In 
the long run the timing variations of both the AXPs and the SGRs are 
much stronger than the timing noise of radio pulsars but comparable 
with the behaviour of accreting sources. For the DTNs, the thermal 
luminosities do not seem compatible with cooling ages and statistics 
if these sources are magnetars. The energy dissipation rates implied by 
the dynamical behaviour of glitching pulsars will provide the luminosities 
of the DTNs under propeller spindown torques supplied by fossil disks on  
neutron stars with conventional 10$^{12}$ G dipole magnetic fields. 
\end{abstract}

\section{The Signature of Large Pulsar Glitches}

A simple model detailed first for the Vela pulsar's glitches 
pictures the glitches as the discrete component of angular 
momentum exchange between the observed crust and other components 
of the neutron star (Alpar et al. 1993 and references therein). 
There is also a continuous part of the
angular momentum exchange between the components. It is believed that 
the interior of the neutron 
star is superfluid, which means that the angular momentum is 
transported by discrete carriers, the quantized vortices of the 
superfluid. There is a simple analogy with an 
electronic circuit that contains resistive as well as capacitive 
elements. The parts of the superfluid where there is a continuous 
vortex current that allows the superfluid to take part in the spindown   
are the resistive elements, where continuous energy dissipation 
accompanies the angular momentum transport. Say these resistive parts 
have total moment of inertia $I_A$. The resistive parts may be disjoint 
regions within which there is a vortex density and vortex current.  
Some of these vortices may be trapped by pinning centers. The resistive 
parts collectively are the analogues of capacitor 'plates' and the 
wires connecting them in an electronic circuit. 
These resistive regions are separated by vortex free regions, analogues 
of the space between capacitor plates. The glitches are 
capacitive discharges of vortices through vortex free regions, say 
of moment of inertia $I_B$, which do not sustain a continuous vortex 
current. 

In a glitch a certain number of vortices get unpinned throughout 
the resistive trap regions A, move through the regions A and 
B rapidly and repin in other parts of the resistive 
trap regions A. This sudden motion of vortices reduces the superfluid 
rotation rate by an amount $\delta\Omega$ throughout the vortex free regions B. 
Under the simple assumption that the area density of the unpinned vortices is 
uniform throughout the resistive regions A, the average decrease in 
superfluid rotation rate throughout the regions A is $\delta\Omega/2$. 
The angular momentum lost by the superfluid through this sudden discharge 
of vortices is gained by the neutron star crust, leading to the 
observed increase in the crust's rotation rate, $\Delta\Omega$, which is 
the glitch:
\begin{equation}
I_c \Delta\Omega = (1/2 \, I_A + I_B) \delta\Omega .
\end{equation}
Here $I_c$ denotes the effective moment of inertia of the crust, 
including all components of the star that are already rigidly 
coupled to the crust on timescales short compared to the rise time 
of the glitch. Application of the model to pulsar glitches 
shows that $I_c$ contains almost the entire moment of inertia 
of the star, with $I_A$ and $I_B$ making up a few per cent 
(Alpar et al. 1993). The physical reason is that the superfluid 
core of the neutron star, which makes up the bulk of the moment 
of inertia, is coupled to the crust strongly, 
on short time scales (Alpar, Langer, \& Sauls 1984). 
The superfluid regions A and B responsible for the glitches 
make up the superfluid in the inner crust of the neutron star, 
which indeed constitutes a few percent of the star's moment of inertia.

The spindown rate of the superfluid, and hence the observed 
spindown rate of the crust, depend on the vortex current that 
achieves the continuous angular momentum transport 
(the internal torque). This vortex current is driven by a 
lag in rotation rate between the superfluid and the crust. 
The lag is offset by the changes in the rotation rates of both 
the crust ($\Delta\Omega$) and the superfluid ($\delta\Omega$) 
at the glitch. There is, therefore, a glitch induced change 
$\Delta\dot{\Omega}$ in the observed spindown rate of the crust. 
Part of this glitch induced change in the spindown rate relaxes 
promptly, as an exponential decay with relaxation times less than a month 
in the Vela pulsar. What is of interest for the 
long term behaviour of the pulsar is a change in the spindown rate 
that does not heal in a prompt exponential decay but instead is observed to 
relax gradually, as a linear function of time,
completing its recovery at the time of the next glitch:
\begin{equation}
\Delta\dot{\Omega} (t) = \Delta\dot{\Omega} (0) (1 - t/t_g ) 
\end{equation}
where t$_g$, the parameter describing the slope of the relaxation in spindown 
(the 'anomalous' second derivative) is found to be
\begin{equation}
t_g = \delta\Omega / |\dot{\Omega}|
\end{equation}
and 
\begin{equation}
\Delta\dot{\Omega} (0) / |\dot{\Omega}| = I_A / I_c .  
\end{equation}
Eqs. (1)-(4) relate the three model parameters, 
$I_A / I_c$, $I_B / I_c$ and $\delta\Omega$, to the observed parameters 
$\Delta\Omega$, $\Delta\dot{\Omega} (0)$ and t$_g$. The interpretation 
of the model is that the glitch induced change in spindown rate 
reflects the complete stopping of vortex current in all resistive regions A, 
because the current is a very sensitive function of the driving lag and stops 
when the lag is offset by the glitch. (This is called 'nonlinear response' 
in the jargon of the model.) The particular form of the recovery 
given in Eqs. (2) and (3) reflects the uniform density of unpinned vortices, as 
does the factor 1/2 in Eq.(1). For the Vela pulsar the timescale 
t$_g$ extracted from the observed second derivative agrees with 
the time to the next glitch within 20 \% . The recovery means that the 
unpinned vortex density is re-pinned by time t$_g$ - so the traps are ready 
to go unstable again on that timescale. Using Eqs. (1)-(4), the long term 
second derivative of $\Omega$ is:
\begin{equation}
\ddot{\Omega} = \frac{I_{A}}{I} \frac{{\dot\Omega}^2}{\delta\Omega}
=(\beta + 1/2)\,[((\Delta\dot\Omega/\dot\Omega)_{-3})^2)/(\Delta\Omega/\Omega)_{-6}] 
\,({\dot\Omega}^2 / \Omega )
\end{equation}
where $\beta = {I_{B}}/{I_{A}}$. This is equivalent to the positive 
anomalous braking index 
\begin{equation}
n \equiv \Omega \ddot{\Omega} / {\dot\Omega}^2 
= (\beta + 1/2)\,((\Delta\dot\Omega/\dot\Omega)_{-3})^2 / (\Delta\Omega/\Omega)_{- 6}.
\end{equation}
The time to the next glitch can be expressed as 
\begin{equation}
t_{g}= 2 \times 10^{-3} \,(\beta + 1/2)^{-1} \, \tau_{sd} \, (\Delta\Omega/\Omega)_{-6}/  
(\Delta\dot\Omega/\dot\Omega)_{-3} 
\end{equation}
where $\tau_{sd} = \Omega/(2|\dot\Omega|)$ is the characteristic 
dipole spindown time.

The implications of this simple pattern in the Vela pulsar's interglitch 
behaviour are:
(i) The neutron star is in a nonlinear response regime, highly sensitive 
to perturbations of the lag. In such a nonlinear response situation the 
energy dissipation rate can be shown to be quite large, 
\begin{equation}
{\dot{E}}_{diss} \sim I_A \omega  |\dot{\Omega}| 
> I_A \delta\Omega  |\dot{\Omega}| \sim 10^{41} |\dot{\Omega}| 
ergs \; s^{-1}.  
\end{equation}
Here $\omega$ is the lag between the rotation rates of the crust and the pinned superfluid. 
$\dot{E}_{diss}$ determines the thermal luminosities of older neutron stars.
(ii) The result, from the fits to the data, that $I_A / I_c$, $I_B / I_c$ are 
$\leq$ 10$^{-2}$ means the core superfluid is strongly coupled to the outer crust 
on short timescales, which in turn implies that precession would be strongly damped. 
The applicability of this simple model to all pulsar glitches and 
interglitch behaviour will imply that points (i) and (ii) are relevant to all neutron stars. 

\section{Evidence for the 'Universality' of Glitch Behaviour ?}

An important caveat in comparing observed glitch parameters with the model is the distinction 
between the total $\Delta\dot\Omega (0)$, which may include contributions 
from transients, and the component of $\Delta\dot\Omega (0)$ associated 
with the long term recovery (Eq.(2)). The observations typically do not resolve this point. 
In some of the glitch observations a single exponentially decaying transient 
fit to postglitch data is presented, sometimes along with a long term second derivative value. 
There may be more than one transient, the relaxation times are expected to be different 
in different pulsars, the data is not uniform and do not cover similar timespans in all 
the different pulsars. From the detailed fits to the Vela pulsar postglitch behaviour, 
it is known that the $\Delta\dot\Omega (0)$ values associated with various 
transients are comparable with that associated with the long term interglitch recovery. 
We adopt the total value of $\Delta\dot\Omega (0)$ as the long term offset in  
$\dot\Omega$ to be used in the model, without subtracting the part 
associated with transient exponential decay that was fitted by the observers to the 
data for some glitches. If the other glitching pulsars indeed behave like the Vela pulsar 
then values of $\beta$ of order one, within an order of magnitude, should be obtained.

Lyne, Shemar, \& Graham-Smith (2000) measured second derivatives of the rotation rate 
from 7 out of 16 glitching pulsars they had observed (excluding the Crab and Vela pulsars). 
Out of these 7 pulsars, those with  quoted relative errors of $>$ 20\% in the second 
derivative, as well as PSR B2224+65 for which the error in $\Delta\dot{\Omega}/\dot{\Omega}$ 
exceeds the nominal value, and PSR B1823-13, for which there is a postglitch second derivative 
measurement, but an observed $\Delta\dot{\Omega}/\dot{\Omega}$ was not reported, were excluded 
from the comparison with the model. The data for the remaining 3 pulsars 
(PSR B1727-33, PSR B1737-30 following the first and fifth glitches and PSR B1800-21) 
were compared with the model, Eqs. (5), (6) and (7). Out of 10 glitching 
southern pulsars (excluding the Vela pulsar) studied by Wang et al. (2000), 
7 have yielded second derivative measurements, with $<$ 20\% quoted errors, after 
an observed glitch. Out of these 7, one pulsar, PSR J1614-5047, is in conflict with 
the model. This pulsar has a negative second derivative 
measurement with quoted errors of only 1.5\% in a timing fit covering a 563 d part 
of its postglitch epoch, after first displaying a positive second derivative over 325 d. 
The quoted second derivative value, however, differs strongly from the plot of 
$\dot\Omega$ data (Wang et al. 2000, Fig.8). For the remaining 6 pulsars (PSR J1048-5832 
following the second and third glitches, 
PSR J1341-6220 following the third and ninth glitches, PSR J1709-4428, PSR J1731-4744, 
PSR J1801-2451 and PSR J1803-2157) the results of Wang et al. (2000) were compared with 
the model.

The 12 glitches from these 9 pulsars form the complete sample, published to date, of 
large glitches with changes in spindown rate at the glitch and with relatively accurate 
post glitch second derivative measurements. Using Eq. (6) values of $\beta$ can be derived 
for each of these glitches. For 10 of the 12 glitches, $\beta$ values of 0.3-16 are obtained. 
These are, within an order of magnitude, similar to the values obtained 
for the Vela pulsar glitches. For the third glitch of PSR J1341-6220 $\beta$ = 43 
while for the glitch observed from PSR J1709-4428 $\beta$ = 1120, clearly in conflict with 
the other $\beta$ values. Eq. (7) gives t$_g$, 
the expected time to the next large glitch for each pulsar. Values of t$_g$ of the 
order of several years are typical. For the pulsars with 
repeated large glitches, $\Delta\Omega/\Omega >$ 10$^{-7}$, 
we can compare the expected and observed 
intervals t$_g$ between these large glitches. The ratio (t$_{g, obs}$ /t$_g$) 
= 1.6 (PSR B1737-30, glitch 1 to glitch 5), 0.96 (PSR J1048-5832, glitch 2 to glitch 3), 
2.5 and 0.7 (PSR J1341, glitch 3 to glitch 5 and glitch 9 to glitch 12). 
It can be concluded, with caution, that the model 
developed for the Vela pulsar may be applicable to most large glitches, and that there may be 
common dynamical behaviour for all neutron stars.
 
Johnston \& Galloway (1999) have obtained braking indices 
for 20 pulsars from which glitches have not been observed. 
Anomalous braking indices were found for all 20 pulsars, with negative 
values in 6 pulsars (in 5 of them with 
relative errors $<$ 20\%) and positive values in the rest (with 
relative errors $<$ 20\% in 8). Johnston \& Galloway (1999) interpreted 
the positive anomalous braking indices 
as due to interglitch recovery, without evoking a specific model. 
For these pulsars, no glitch has occurred during the timespan of the observations. 
The values of the parameter $\beta$ obtained 
from Eq.(6) with the measured braking indices and with nominal 
glitch values $\Delta\Omega/\Omega$ = 10$^{-7}$ 
and $\Delta\dot\Omega/\dot\Omega$ = 10$^{-3}$ 
are of order $\beta \sim 0.36 - 4.5$, 
consistent with all these pulsars having similar 
glitches with similar ratios of the vortex trap (capacitive) 
and vortex creep (resistive) moments of inertia. The negative braking indices were 
interpreted by Johnston \& Galloway (1999) as reflecting an unresolved 
glitch during their observation time spans. Since all glitches involve a
negative step (an increase in the absolute value) of the rate of spindown, 
and since the pulsars were not monitored continuously, a glitch occurring 
between two timing observations would lead to a negative $\ddot\Omega$ 
estimate, equivalent to a negative braking index. The fractional changes in 
the spindown rate in the four unobserved glitches are inferred from the negative 
braking indices of Johnston \& Galloway (1999) according to:
\begin{equation}
\Delta\dot{\Omega}_i \, = \, \ddot{\Omega}_i \, t_i.
\end{equation} 
These values of $\Delta\dot{\Omega}/\dot{\Omega} \sim $ 5 $\times$ 10$^{-4}$ - 4 $\times$ 
10$^{-3}$ are typical for glitching pulsars. 
The probability that 4 out of the 20 pulsars in the Johnston \& Galloway (1999) sample 
have had unresolved glitches within the observation timespans, so that they have negative 
anomalous second derivatives, can be evaluated using the observation timespans devoted to 
each pulsar and model estimates for the interglitch time intervals t$_g$ of each pulsar, using 
Eq. (7). This probability turns out to be as large as 0.16 for one particular scaling of 
t$_g$ with pulsar rotation frequency. Thus positive and negative braking index measurements 
from pulsars that have not been observed to glitch are also consistent with the model, on 
the account of $\beta$ and $\Delta\dot{\Omega}$ values inferred and also statistically. 

\section{SGR and AXP Timing Behaviour: Comparison with Isolated Pulsars}
 
The timing behaviour of AXPs and SGRs could provide a clue as to 
whether these objects are magnetars or accreting sources. If magnetars are not 
qualitatively different from the radio pulsars, they should have very steady 
spindown rates, with a low level of timing noise, like the radio pulsars. 
All radio pulsars show steady spindown rates for as long as they have been observed, 
with timing noise levels (Arzoumanian et al. 1994, Alpar, Nandkumar, \& Pines 1986) that are 
significantly lower than those of the accreting sources (Bildsten et al. 1997). 
Kaspi, Chakrabarty \& Steinberger (1999) reported such quiet spindown in RXTE timing 
observations of the AXPs 1RXS J170849.0-400910 (for a period of 1.4 yr) and 
1E 2259+586 (for 2.6 yr). In earlier observations the latter source, 
and a third AXP, 1E 1048.1-5937, had been observed to display noisy spindown behaviour, 
similar to torque noise typical of accreting sources (Mereghetti 1995, Baykal \& Swank 1996, 
Corbet \& Mihara 1997). Baykal et al. (2000) showed from later archival RXTE 
observations that 1E 1048.1-5937 continued to display noisy torque behaviour while 
1E 2259+586 had a stable luminosity behaviour, a luminosity time series with a 
particularly low level of fluctuations, during the same epoch when Kaspi et al. (1999) 
had observed quiet spindown. Not all 
AXPs have been observed in epochs of quiet spindown; those that do have extended quiet 
spindown episodes have been observed to have strong timing noise at other epochs; and 
quiet spindown (low torque noise) coincides with low luminosity noise. Thus the observed timing 
behaviour of AXPs is consistent with an accretion scenario. Furthermore, 
Baykal et al (2001) show, by adding recent RXTE pulse period measurements to earlier 
timing results, that a well known accreting source, the high mass X-ray binary 
4U 1907+09, has been spinning down at an almost constant rate, with very low level timing noise,
for almost 15 years, clinching the case that quiet spindown does not necessarily imply 
a non-accreting source. To conclude that a source is like a radio pulsar in spindown behaviour,
it would have to be spinning down with low noise all the time.    

The extended quiet spindown of the AXP 1RXS J170849.0-400910, continuing for more than two years 
in frequently monitored RXTE timing, was interrupted by a sudden glitch (Kaspi, Lackey, \& 
Chakrabarty 2000). This glitch seems to be a typical example of the universal large pulsar 
glitches described above in both the fractional change in frequency, 
$\Delta\Omega/\Omega \cong 6.2 \times$ 10$^{-7}$, and in the fractional change 
in spindown rate, $\Delta\dot{\Omega}/\dot{\Omega} \cong 3.8 \times$ 10$^{-3}$. 
This is the first instance of the 'universal' glitch occurring in a neutron star 
other than a radio pulsar and shows that the same neutron star dynamics shows 
up in response to spindown under an external torque provided the spindown proceeds 
at a constant rate. Data is not yet available for checking the second derivative 
characteristic of the interglitch behaviour observed in radio pulsars. The glitch model, 
Eq.(7), can be used to infer that t$_g \cong$ 1 yr, so that there is an appreciable 
probability of observing such a glitch within the two year timespan of the observed 
quiet spindown. Indeed if the quiet spindown continues, repeated glitches should be expected. 

The possibility that the observed timing residuals of AXPs during noisy periods 
arise from a series of unresolved glitches can be explored with various combinations of model 
glitches that fall in gaps in the observations ( Heyl \& Hernquist 1999). The required event 
sizes are $\Delta\Omega/\Omega \sim $ 10$^{-4}$ for the AXP 1E 1048.1-5937 
(much larger than the values for the radio 
pulsar glitches) and $\Delta\Omega/\Omega \sim $ 10$^{-6}$ 
for the AXP 1E 2259+586. Taking $\Delta\dot\Omega/\dot\Omega \sim $ 10$^{-3}$,  
and $\beta$ as large as 10, Eq. (7) predicts average event intervals 
t$_g \sim$ 10$^2$ yrs or longer for both AXPs
while the interval required to explain the timing excursions in the AXPs 
is a few years in some fits. Thus both the size and the rate of the hypothesized 
unresolved glitches in SGRs are large compared to the pulsars. 

Woods et al. (2001) have recently reported all timing observations, 
through January 2001, of the two SGRs with $\dot\Omega$ 
measurements, SGR 1900+14 and SGR 1806-20. No events resembling pulsar glitches,  
i.e. discrete and sudden spin-up events against the background of spindown,  
were either resolved or implied by mismatches in extrapolations of timing 
solutions into gaps between observations. In both sources the 
most prominent characteristic of the timing behaviour is changes in the 
spindown rate by factors of as much as 3-4 extending over periods 
of several months. These variations were followed through frequently 
sampled RXTE-PCA observations in 2000. Such large variations cannot 
be due to changes in moment of inertia of the crust 
through quakes or gradual plastic deformation. If the universal behaviour 
of radio pulsar glitches is taken as a guide to the internal torques, 
and the distribution of moment of inertia in various components of 
the neutron star, then the limitation on the crust core 
coupling time, $\tau <$ 450 P(s), obtained by extrapolating the limit 
$\tau <$ 40 s for the Vela pulsar glitches (Dodson, Lewis, \& McCulloch 2001) 
through the theoretical explanation of the tight crust-core coupling 
(Alpar et al. 1984), implies that during the large changes 
of $\dot\Omega$ the entire star is rotating rigidly with the observed 
crust. Thus a change by a factor of 3-4 in the spindown rate, if due to 
a structural change, would imply a corresponding change in the moment of 
inertia of the entire neutron star. 
The variations in spindown rate therefore are likely to reflect variations 
in the external torque on the star. Such changes are easy to understand if 
the external torque is an accretion torque. 
Changes in spindown (or spinup) rate by such factors of 3-4 is 
common in accreting sources. The overall power in timing noise is 
reported by Woods et al. (2001) to be compatible with the range of noise 
strengths in accreting sources and much stronger than the timing noise 
in radio pulsars. If the external torque is a dipole spindown torque 
with a magnetar field then the strength of the variations in the 
spindown rate (timing noise) being so much larger than timing noise strength 
in the radio pulsars must be ascribed to a qualitative change in the external 
torque when the magnetic field is in the 10$^{14}$ G magnetar range rather 
than the 10$^{12}$ G range of the radio pulsars. Such a qualitative change in 
torque behaviour must occur at surface magnetic fields larger than B$_0$ = 1.1 $\times$ 
10$^{14}$ G at the 
neutron star's magnetic pole, since the radio pulsar with the largest inferred 
field value, PSR J1814-1744 (Camilo et al. 2000) with B$_0$ = 1.1 $\times$ 
10$^{14}$ G, does not exhibit the large torque variations 
observed from the SGRs. But then neither do the highest magnetic field radio pulsars 
exhibit soft gamma ray bursts. In the magnetar model, the gamma ray burst behaviour 
and the torque behaviour of neutron stars as isolated rotating dipoles must go 
through a transition at B$_0 >$ 1.1 $\times$ 10$^{14}$ G.

Are the SGRs similar to the AXPs in their overall timing behaviour? It may well be that 
the SGRs also go through extended quiet spindown phases that have been observed from 
some AXPs and from the known accreting source 
4U 1907+09 but that for the SGRs the quiet spindown phases have not yet occurred during 
the era covered by observations. In any case the SGRs do have the noisy timing behaviour 
characteristic of the accreting sources. It is the occurrence of the repeated soft 
gamma ray bursts, the defining characteristic of the 
SGRs, and not the timing behaviour during the quiescent X-ray phases, that sets the 
SGRs apart from their potential relatives the AXPs, and from the accreting X-ray sources. 
The recent observations of Woods et al. (2001) most significantly make 
the point that no soft gamma ray bursts took place in either SGR 1900+14 or SGR 1806-20 during 
the changes in spindown rate that were followed through the RXTE PCA observations in 2000, 
which, as argued above, probably reflect 
changes in the external torque. During burst active epochs, in 
1998-1999 for SGR 1900+14 and late 1996-early 2000 for SGR 1806-20, sparsely sampled timing 
data imply a steady long term spindown rate. In any case, models for the bursts must confront 
why the largest observed variations in external torque are separated from any burst activity,
by many years in SGR 1900+14 and by about 10 months in SGR 1806-20, as Woods et al. (2001) find.

\section{Thermal Luminosities of DTNs: Cooling Magnetars or Energy Dissipation?} 

The dim thermal neutron stars discovered in ROSAT surveys (Treves et al. 2000) have similarities 
with the AXPs and SGRs. In particular the three dim thermal neutron stars with measured 
periods all have periods similar to the AXP and SGR periods (P = 8.37 s, 5.16 s and 22.7 s, 
respectively, for RXJ 0720.4-3125, RBS 1223 and RXJ 0420.0-5022). Distance estimates for the 
DTNs are typically in the range of 100 pc, with large uncertainties. With n=6 detected 
nearby examples, the DTNs must form a very abundant population in the galactic plane where 
they all lie. Scaling to the entire galactic plane with a radius of 10 kpc, and assuming 
a galactic birthrate R = R$_{-2} \times$ 10$^{-2}$ yr$^{-1}$ gives the relation
\begin{equation}
\tau_6 = n ( R_{-2})^{-1} (d/100 pc)^{-2}    
\end{equation}
between the lifetime $\tau_6$ ( in 10$^6$ yrs) of the population with n objects detected 
out to distance d. 

Like the AXPs and SGRs the DTNs have been proposed as examples of magnetars. This is based 
on the interpretation of their X-ray luminosity as cooling luminosity, which implies that 
the neutron star is young. The slow rotation rate in conjunction with young age then implies 
efficient braking, ie that a magnetic field of magnetar strength must have spun down the 
neutron star through magnetic dipole radiation. According to the whole range of standard 
cooling calculations, the luminosity of a young neutron star, of 
age 10$^2$ yrs$ < \tau <$ 10$^5$ yrs, is in the range L$_x \sim$10$^{33}$ 
erg s$^{-1}$ - 10$^{34}$ erg s$^{-1}$ (see, for example, Fig.5 of \"Ogelman (1995)). 
The cooling luminosity is about 10$^{32}$ erg s$^{-1}$ at $\tau = $10$^6$ yrs and drops to 
about 10$^{26}$ erg s$^{-1}$ by $\tau $= 3$\times$10$^6$ yrs .

For RBS 1223, comparison of periods determined from recent Chandra observations and 
earlier ROSAT HRI observations lead to an estimate of the spindown rate, 
$\dot\Omega = -(1.7 - 4.7) \times 10^{-12}$ (Hambaryan et al. 2001), similar to the spindown 
rates of AXPs and SGRs. The uncertainty in the spindown rate reflects the uncertainties in 
the two period determinations and is not a measurement of timing excursions. 
If the source is a rotating dipole, the magnetic moment in units of 10$^{30}$ G is 
$\mu_{30} \cong (I_{45} P \dot{P}_{-15})^{1/2}$ = 190-320, implying dipole fields in 
the magnetar range. For this source the dipole spindown time $\tau = P /2\dot{P}$ 
is found to be 4000-12000 yrs. Distance estimates give d=100-200 pc or d= 700-1500 pc. 
Only for the latter case, at the largest distance d=1500 pc, the luminosity 
L$_x \cong$ 10$^{33}$ erg s$^{-1}$ of RBS 1223 is barely consistent with the lowest 
luminosities allowed by standard cooling models at the estimated age of $\sim$ 10$^4$ yrs. 

For another DTN, RXJ 1856.6-3754, there is a kinematic age determination of 10$^6$ yrs based on 
proper motion measurements and on an inferred birthplace 
in the Upper Sco OB association (Walter 2001). Differences in parallax measurements with HST 
are likely to be resolved with new HST data. If this source has a period in the 
range 5 s - 23 s of the other DTNs, and is spinning down by magnetic dipole radiation, 
the magnetic moment implied by taking the kinematic age as the dipole spindown time is 
\begin{equation}
\mu_{30} \cong 4 P (\tau_6 / I_{45})^{-1/2}
\end{equation}
giving $\mu_{30}$ = 20-92.
Adopting the distance value of 140$\pm$40 pc (Kaplan, van Kerkwijk \& 
Anderson 2001) gives a thermal X-ray luminosity L$_x \sim $ 10$^{32}$ erg/s, which 
is consistent with standard cooling at the estimated magnetar age.

The two other DTNs with measured periods, RXJ 0720.4-3125 (P=8.37s) and RXJ 0420.0-5022 
(P=22.7s) have spindown ages of 1.1 $\times$ 10$^5$ yrs and 8.2 $\times$ 10$^5$ yrs 
respectively, if they are magnetars with $\mu_{30}$ = 100. With their observed thermal 
X-ray fluxes, 
both sources must be at distances of 500-1000 pc if the luminosities are standard cooling 
luminosities. If these are the distances to which the observed sample of DTNs extends, 
then from Eq.(10), the lifetime of DTNs is $\tau$ = (6-25) $\times$ 10$^4 (R_{-2})^{-1}$ yrs. 
Of the four DTNs with measured periods and age determinations or estimates 
(within the magnetar model, except for RXJ 1856.6-3754 which has a kinematic age estimate), 
only one, RXJ 0720.4-3125, has an age estimate within this range 
of the statistically determined lifetimes. 
More and better distance and age determinations will resolve whether this difficulty for 
the magnetar model is real. 

The evidence, from pulsar glitches, of a common dynamical behaviour for all pulsars, implies a 
different origin for the X-ray luminosity, energy dissipation in the exchange of angular 
momentum between the components of the neutron star rotating at different rates. The observed 
postglitch and interglitch behaviour indicate strong, nonlinear coupling between the 
components. This implies that the energy dissipation rate, which is 
proportional to the lag $\omega$ between the crust and 
the pinned superfluid components is large. The lag must be greater than the reduction 
$\delta\Omega$ in the rotation rate of the pinned superfluid at a glitch. With the value of 
$\delta\Omega \sim$ 10$^{-2}$ rad s$^{-1}$ and of neutron star crust moments of inertia 
commonly inferred from glitches, a lower bound on the energy dissipation rate, 
given in Eq.(8) is obtained. Using this together with an upper bound obtained from 
thermal X-ray observations of radio pulsars (Alpar et al. 1987, Yancopoulos, Hamilton, 
\& Helfand 1994), the observed luminosities of the DTNs, 
interpreted as due to energy dissipation in the neutron star yield bounds on the 
spindown rate (Alpar 2001):
\begin{equation}
L_x / 10^{43} < |\dot{\Omega}| < L_x / 10^{41} . 
\end{equation}
The range of spindown rates inferred from DTN fluxes, at the more likely distances of 
the order of 100 pc, are $|\dot{\Omega}| \sim 10^{-13}-10^{-10}$ rad s$^{-2}$. This range 
includes the observed spindown rate of RBS 1223 and overlaps with the range of spindown rates 
of the AXPs and SGRs. The DTN spindown can be understood as due to propeller torques 
from a fossil accretion disk on a neutron star with conventional magnetic moment, 
$\mu \sim $10$^{30}$ G (Alpar 2001). In the propeller epoch, the lifetime of a neutron 
star is not bounded by the duration of initial cooling, as it is in the case of the 
cooling magnetar model. Lifetimes of a few 10$^6$ yrs statistically inferred 
from the likely distances of the order of 100 pc are quite acceptible. 
The requirement of thermal luminosities for old neutron stars, arising from large 
energy dissipation rates is a strong consequence of the observed glitch behaviour if it 
prevails universally among all neutron stars.

\acknowledgments

This work was supported by the Astrophysics and Space Forum at Sabanc{\i} University, by 
T\"UB\.{I}TAK- \c{C}G-4 and by the Turkish Academy of Sciences.

\end{document}